\documentclass[aps,preprintnumbers,floatfix,letterpaper,superscriptaddress,footinbib,twocolumn,tightenlines]{revtex4-1}
\pdfoutput=1
\usepackage[utf8x]{inputenc}

\usepackage{amsmath,amssymb,amsthm}

\usepackage{bm}       
\usepackage{graphicx} 
\usepackage{epsfig}   
\usepackage[usenames]{color}
\usepackage{tabularx}

\DeclareMathOperator{\erf}{erf}

\begin{document}

\title{Finite-key analysis of high-dimensional time-energy entanglement-based quantum key distribution}

\author{Catherine Lee}
\affiliation{Research Laboratory of Electronics, Massachusetts Institute of Technology, Cambridge, Massachusetts 02139, USA}
\affiliation{Department of Physics, Columbia University, New York, New York 10027, USA}

\author{Jacob Mower}
\affiliation{Research Laboratory of Electronics, Massachusetts Institute of Technology, Cambridge, Massachusetts 02139, USA}
\affiliation{Department of Electrical Engineering, Columbia University, New York, New York 10027, USA}

\author{Zheshen Zhang}
\affiliation{Research Laboratory of Electronics, Massachusetts Institute of Technology, Cambridge, Massachusetts 02139, USA}

\author{Jeffrey H. Shapiro}
\affiliation{Research Laboratory of Electronics, Massachusetts Institute of Technology, Cambridge, Massachusetts 02139, USA}

\author{Dirk Englund}
\affiliation{Research Laboratory of Electronics, Massachusetts Institute of Technology, Cambridge, Massachusetts 02139, USA}
\affiliation{Department of Electrical Engineering, Columbia University, New York, New York 10027, USA}
\affiliation{Department of Applied Physics and Applied Mathematics, Columbia University, New York, New York 10027, USA}

\date{\today}

\begin{abstract}
We present a security analysis against collective attacks for the recently proposed time-energy entanglement-based quantum key distribution protocol, given the practical constraints of single photon detector efficiency, channel loss, and finite-key considerations. 
We find a positive secure-key capacity when the key length increases beyond $10^4$ for eight-dimensional systems. The minimum key length required is reduced by the ability to post-select on coincident single-photon detection events. 
Including finite-key effects, we show the ability to establish a shared secret key over a 200 km fiber link.
\end{abstract}

\maketitle

\section{Introduction}

High-dimensional quantum key distribution (QKD) \cite{PRA.61.062308} allows two parties, Alice and Bob, to establish a secret key at a potentially higher rate than that afforded by two-level QKD protocols \cite{BB84,E91}. 
When the photonic states span a high-dimensional Hilbert space, more than one bit of secure information can be shared per single photon detected. 
Additionally, increasing the dimension of a QKD protocol can improve resilience to noise \cite{PRL.88.127902}. 
High-dimensional QKD protocols have been implemented by encoding information in various photonic degrees of freedom, including position-momentum \cite{PRL.100.110504}, time \cite{PRL.84.4737,PRL.93.010503,PRL.98.060503,PRA.66.062304,OptLett.31.2795,2013.PRA.Mower.do-qkd}, and orbital angular momentum  \cite{Nature.412.313,PRL.89.240401,PRL.92.167903,PRA.88.032305}.

\begin{figure}
\begin{center} 
    \includegraphics[scale=0.37]{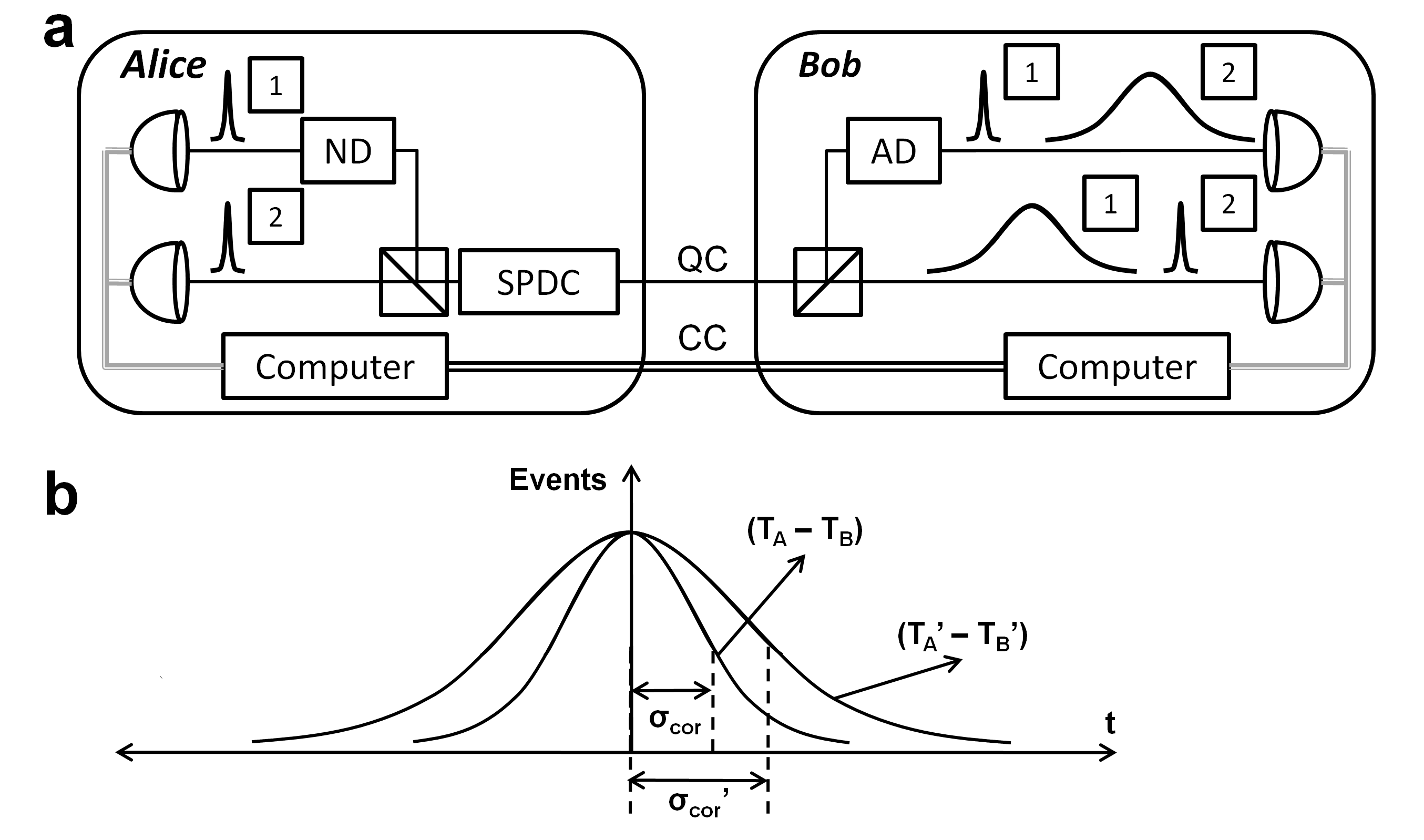}
    \caption{a) Schematic of the DO-QKD setup. Alice holds the SPDC source, keeps one photon, and sends the other to Bob. In case 1, Alice measures in the dispersed arrival-time basis, and in case 2, she measures in the arrival-time basis. Bob must measure in the same basis as Alice for their measurements to be correlated. QC is quantum communication, CC is classical communication, ND is normal dispersion, and AD is anomalous dispersion. b) Depiction of decreased photon correlations measured by Alice and Bob, from the ideal correlation time $\sigma_\mathrm{cor}$ to the observed $\sigma_\mathrm{cor}'$.
    }
    \label{setup}
\end{center}
\end{figure}

Here, we consider the recently proposed dispersive-optics QKD protocol (DO-QKD), which employs energy-time entanglement of pairs of photons. 
We recently proved security against collective attacks for this protocol in the limit of infinite key length \cite{2013.PRA.Mower.do-qkd}. 
In DO-QKD, the photon pairs are generated by a spontaneous parametric downconversion (SPDC) source held by Alice. The largest possible dimension $d$ of the protocol is given by the Schmidt number, i.e., the number of possible information eigenstates in the system. 
This is approximately $d \equiv \sigma_\mathrm{coh}/\sigma_\mathrm{cor}$ \cite{PRL.92.127903,PRL.98.060503}, 
where $\sigma_\mathrm{coh}$ is the coherence time of the SPDC pump field, and $\sigma_\mathrm{cor}$ is the correlation time between photons, which is set by the phase-matching bandwidth of the SPDC source. 
Alice keeps one photon and sends the other to Bob, and shared information is generated from the correlated photon arrival times measured on single photon detectors by Alice and Bob. 
Figure~\ref{setup}a presents a schematic of the setup.

In DO-QKD, conjugate measurement bases are implemented using group velocity dispersion (GVD). If Alice applies normal dispersion and Bob applies anomalous dispersion, the original correlations between their photons can be recovered \cite{PRA.45.3126}. 
However, if only one party applies dispersion and the other does not, the timing correlations are lost or severely diminished, depending on the magnitude of the dispersion applied. 
Alice and Bob randomly choose to apply---or not apply---dispersion to their photons before measuring their arrival times. Measurements made without dispersion are referred to as being in the arrival-time basis and measurements made with dispersion are in the dispersed arrival-time basis. 
After the measurement stage of the protocol, Alice and Bob communicate their basis choices and keep only the data from time frames in which they each registered a single detection event while using the same basis. They publicly compare a subset of their raw keys to bound an eavesdropper's shared information.
Then they use error correction and privacy amplification \cite{PRL.77.2818} to extract identical secret keys.

The security analysis of DO-QKD relies on estimating the covariance matrix of Alice's and Bob's measurements. Specifically, Alice and Bob use their publicly compared raw key to estimate the increase in the correlation time of their photons from $\sigma_\mathrm{cor}$ 
to $\sigma_\mathrm{cor}'$, the experimentally observed correlation time (see Figure~\ref{setup}b). 
The precision of Alice and Bob's estimation of $\sigma_\mathrm{cor}'$ increases with the sample size; however, publicly comparing a greater fraction of their measurements reduces the amount of raw key that can be used to generate the secret key. In practice, Alice and Bob have a finite number of measurements, so they must find the optimal compromise between the conflicting goals of accurately estimating parameters and maximizing the length of their secret key. 

Alice and Bob's finite number of measurements also requires a generalization of our previous security proof for DO-QKD \cite{2013.PRA.Mower.do-qkd}, which relied on the asymptotic limit: Alice and Bob's keys and the data subset used for parameter estimation were assumed to be infinitely long. 
Here, we extend our previous security analysis to show that DO-QKD is secure against collective attacks,
given the practical constraints of single-photon detector efficiency, channel loss, and finite-key considerations \cite{PRL.100.200501,PRA.82.030301,PRA.83.039901,NewJPhys.11.045024,NewJPhys.12.123019,PRA.81.062343,PRL.109.100502,NComm.3.634,PRL.110.030502}.

\section{Finite-key analysis for arbitrary basis selection probabilities}
\subsection{Asymmetric basis selection}

In the standard QKD protocols \cite{BB84,E91,RevModPhys.74.145}, 
Alice and Bob selected between the two measurement bases with equal probabilities, limiting the probability of generating a shared character of key to at most 50\%. It was later suggested \cite{JCryptology.18.133} that the efficiency of a QKD protocol could be increased asymptotically to 100\% if Alice and Bob choose one measurement basis with a greater probability than the other, which increases the likelihood that Alice and Bob will make measurements in the same basis. 
We will take the same approach here.

Without further modification to our protocol, Eve could exploit Alice and Bob's asymmetric selection. If Eve were aware of Alice and Bob's basis choice probabilities, then by using only the dominant basis, she could eavesdrop while introducing fewer timing errors in the conjugate basis, i.e., a smaller observed increase in the correlation time. If Eve chooses to eavesdrop in the arrival-time basis, she would introduce more errors in the dispersed-arrival-time basis. This gives Eve a better chance of remaining undetected by Alice and Bob. 
To prevent this possibility, Alice and Bob must further modify their protocol: they divide their data according to the measurement basis used, and they estimate parameters, such as the correlation time, separately for each basis.

When implementing DO-QKD using asymmetric basis selection, we assume that Alice and Bob choose to measure in the arrival-time basis (in which photons are measured directly without dispersion) with probability $p > 1/2$; that is, Alice and Bob apply GVD to fewer than half of the signal photons. 
The exact value of $p$ must then be chosen, along with other parameters, to optimize the secure-key capacity for a given finite number of measurements, as described below.

\subsection{Finite-key effects on secure-key capacity}

Outside the asymptotic limit,a protocol can be only $\varepsilon_s$-secure, where $\varepsilon_s$ is the tolerated failure probability of the entire protocol \cite{PRL.100.200501}. 
The entire protocol is said to fail if, at its conclusion, unbeknownst to Alice and Bob, the eavesdropper holds information about their secret key. 
The security parameter $\varepsilon_s$ is the sum of the failure probabilities of each stage of the protocol: 
\begin{equation}
 \varepsilon_s = \varepsilon_{EC} + \varepsilon_{PA} + \varepsilon_{PE} + \bar{\varepsilon},
 \label{secParam}
\end{equation}
where $\varepsilon_{EC/PA/PE}$ are the probabilities that error correction, privacy amplification, or parameter estimation, respectively, fail \cite{NewJPhys.12.123019}.
Error correction fails if Alice and Bob are unable to obtain identical keys. Privacy amplification fails if it leaks information to the eavesdropper. Parameter estimation fails if the real parameter lies outside of the confidence interval set by $\varepsilon_{PE}$. 
The $\bar{\varepsilon}$ term in \eqref{secParam} accounts for the accuracy of estimating the smooth min-entropy, which characterizes the amount of secure information that can be extracted using privacy amplification \cite{PRL.100.200501}. 
Failure of any stage of the protocol implies that Alice and Bob are unaware that something has gone wrong \cite{PRA.81.062343}.

The finite-key secure-key capacity for the DO-QKD protocol can then be written as \cite{PRL.100.200501,PRA.82.030301,PRA.83.039901,NewJPhys.11.045024,NewJPhys.12.123019,PRA.81.062343}:
\begin{equation}
\begin{split}
 r_N = &\frac{n}{N}\left( r_{DO} - \frac{1}{n}\log_2 \frac{2}{\varepsilon_{EC}} - \frac{2}{n}\log_2 \frac{1}{\varepsilon_{PA}} \right. \\ & \left. - (2 \log_2 d + 3)\sqrt{\frac{\log_2 (2/\bar\varepsilon)}{n}} \right).
\end{split}
 \label{ssRate}
\end{equation}
Here $r_{DO}$ is the secure-key capacity in the asymptotic regime, which was derived in Ref.~\cite{2013.PRA.Mower.do-qkd}. The units in \eqref{ssRate} are bits per coincidence (bpc), i.e., bits per frame in which Alice and Bob each detect only one event. 
$N$ is the number of instances in which Alice and Bob both detect a single photon in a measurement frame. The parameter $n = p^2N$ denotes the number of frames in which Alice and Bob both chose the arrival-time basis, where $p$ is the probability that the arrival-time basis is chosen. We assume that Alice and Bob use the same value of $p$. The subtracted terms on the right-hand side of Eq.~\eqref{ssRate} represent the corrections to $r_{DO}$ due to the finite key length. 

The factor $n/N$ in \eqref{ssRate} reflects the fact that not all of the coincidences detected by Alice and Bob contribute to key generation because some coincidences must be sacrificed for parameter estimation. In particular, we assume that all $m = (1-p)^2N$ coincidences in the dispersed arrival-time basis are used for parameter estimation. Alice and Bob also sacrifice $m$ of the coincidences in the arrival-time basis to estimate parameters for that basis, leaving $n-m$ coincidences in the arrival-time basis for key generation.

For each value of $N$, we maximize $r_N$ by optimizing the parameter set \{$\varepsilon_{PA}$, $\varepsilon_{PE}$, $\bar{\varepsilon}$, $p$\}; thus the basis choice probability $p$ is a function of $N$, the number of signals exchanged. The security parameter $\varepsilon_s$ is determined beforehand by Alice and Bob's security requirements, and $\varepsilon_{EC}$ is fixed by the choice of error correction code.
Additionally, the calculation of $r_{DO}$ must be modified to include the effects of finite key length on parameter estimation.

\subsection{Modified asymptotic secure-key capacity and parameter estimation}

The asymptotic secure-key capacity $r_{DO}$ is given by \cite{2013.PRA.Mower.do-qkd}:
\begin{equation}
 r_{DO} = \beta I(A;B) - \chi(A;E),
\end{equation}
where $\beta$ is the reconciliation efficiency, $I(A;B)$ is Alice and Bob's Shannon information, and $\chi(A;E)$ is Alice and Eve's Holevo information.
Since Alice and Bob use only measurements made in the arrival-time basis for the key, their Shannon information is calculated using only the contribution from the arrival-time basis. This calculation includes the effects of detection efficiency, timing jitter, and dark counts. 
To calculate the Holevo information, Alice and Bob must determine the covariance matrix of their data. To do this, they must estimate the increase in their photons' correlation time from $\sigma_\mathrm{cor}$ to $\sigma_\mathrm{cor}'$, as depicted in Figure~\ref{setup}b. 

The covariance matrix $\Gamma$ is given by
\begin{equation}
 \Gamma = \begin{pmatrix}
           \gamma_{AA} & (1-\eta)\gamma_{AB}\\
 (1-\eta)\gamma_{BA} & (1+\epsilon)\gamma_{BB}
          \end{pmatrix},
\end{equation}
where $\Gamma$ is a four-by-four matrix composed of four two-by-two submatrices. Each submatrix $\gamma_{JK}$ for $J,K = A,B$ describes the covariance between the measurements of parties $J$ and $K$. The submatrices are given by
\begin{eqnarray*}
  \gamma_{AA} &=& \begin{pmatrix}
           \frac{u+v}{16} & -\frac{u+v}{8k}\\
 -\frac{u+v}{8k} & \frac{(u+v)(4k^2+uv)}{4k^2uv}
          \end{pmatrix}, \\
          \gamma_{AB} &=& \gamma_{BA}^T = \begin{pmatrix}
           \frac{u-v}{16} & \frac{u-v}{8k}\\
 -\frac{u-v}{8k} & -\frac{(u-v)(4k^2+uv)}{4k^2uv}
          \end{pmatrix}, \\
\gamma_{BB} &=& \begin{pmatrix}
           \frac{u+v}{16} & \frac{u+v}{8k}\\
 \frac{u+v}{8k} & \frac{(u+v)(4k^2+uv)}{4k^2uv}
          \end{pmatrix},
\end{eqnarray*}
where $u = 16\sigma_\mathrm{coh}^2$ and $v = 4\sigma_\mathrm{cor}^2$ \cite{2013.PRA.Mower.do-qkd}. 
In $\Gamma$, $\eta$ represents the decrease in correlations, and $\epsilon$ represents the excess noise. These two parameters quantify the effects of an eavesdropper, channel noise, and setup imperfections. Without loss of generality, we assume that $\eta$ and $\epsilon$ are the same for both bases.

Alice and Bob can obtain values for $\eta$ and $\epsilon$ using their estimate for $\sigma_\mathrm{cor}'$. We define the parameter $\xi$, which quantifies the increase in the correlation time: $\sigma_\mathrm{cor}'^2 = (1+\xi) \sigma_\mathrm{cor}^2$.
Then, the relationship between $\eta$, $\epsilon$, and $\xi$ is given by 
\begin{equation}
 \epsilon = \frac{-2\eta(d^2-\frac{1}{4})+\xi}{d^2+\frac{1}{4}}.
 \label{constraint}
\end{equation}
Alice and Bob estimate $\xi$ from their data and choose values of $\eta$ and $\epsilon$ that maximize the Holevo information (thereby minimizing $r_{DO}$) and satisfy Eq.~\eqref{constraint} and the following conditions \cite{2013.PRA.Mower.do-qkd}: (i) Eve cannot increase Alice and Bob's Shannon information by interacting with only Bob's photons due to the data processing inequality; (ii) the symplectic eigenvalues of the covariance matrix are greater than $\frac{1}{2}$ such that the Heisenberg uncertainty relation is satisfied; (iii) Eve can only degrade (and not improve) Alice and Bob's measured arrival-time correlation.

Alice and Bob sample only part of their data to estimate $\sigma_\mathrm{cor}'$. In the finite-key regime, it is important to know how well their estimate represents the entire dataset. 
Because Alice and Bob's arrival times, $T_A$ and $T_B$, in a post-selected frame are jointly-Gaussian random variables, and the sequence of these measurements are statistically independent, their estimate for $\sigma_\mathrm{cor}'$, denoted $\hat{\sigma}_\mathrm{cor}'$, has a $\chi^2$ distribution: 
\begin{equation}
(m-1)\frac{\hat{\sigma}_\mathrm{cor}'^2}{\sigma_\mathrm{cor}^2} \sim  \chi^2(1-\varepsilon_{PE},m-1).
\end{equation}
An upper bound on $\sigma_\mathrm{cor}'$ is then given by \cite{PRA.81.062343}: 
\begin{equation}
 (\sigma_\mathrm{cor,max}')^2 = \sigma_\mathrm{cor}^2 + \frac{2}{\sqrt{m}}\erf^{-1}(1-\varepsilon_{PE})\hat{\sigma}_\mathrm{cor}'^2.
\end{equation}
This bound is valid for the confidence interval $1-\varepsilon_{PE}$. 
Then, the largest possible estimate for $\xi$ within the confidence interval is
\begin{equation}
 \xi_\mathrm{max} = \frac{(\sigma_\mathrm{cor,max}')^2}{\sigma_\mathrm{cor}^2}-1.
\end{equation}
Now, Alice and Bob can use their estimate for $\xi_\mathrm{max}$ to calculate the most pessimistic secure-key capacity.

\section{Numerical results}

\begin{figure}
\begin{center}
    \includegraphics[width=\linewidth]{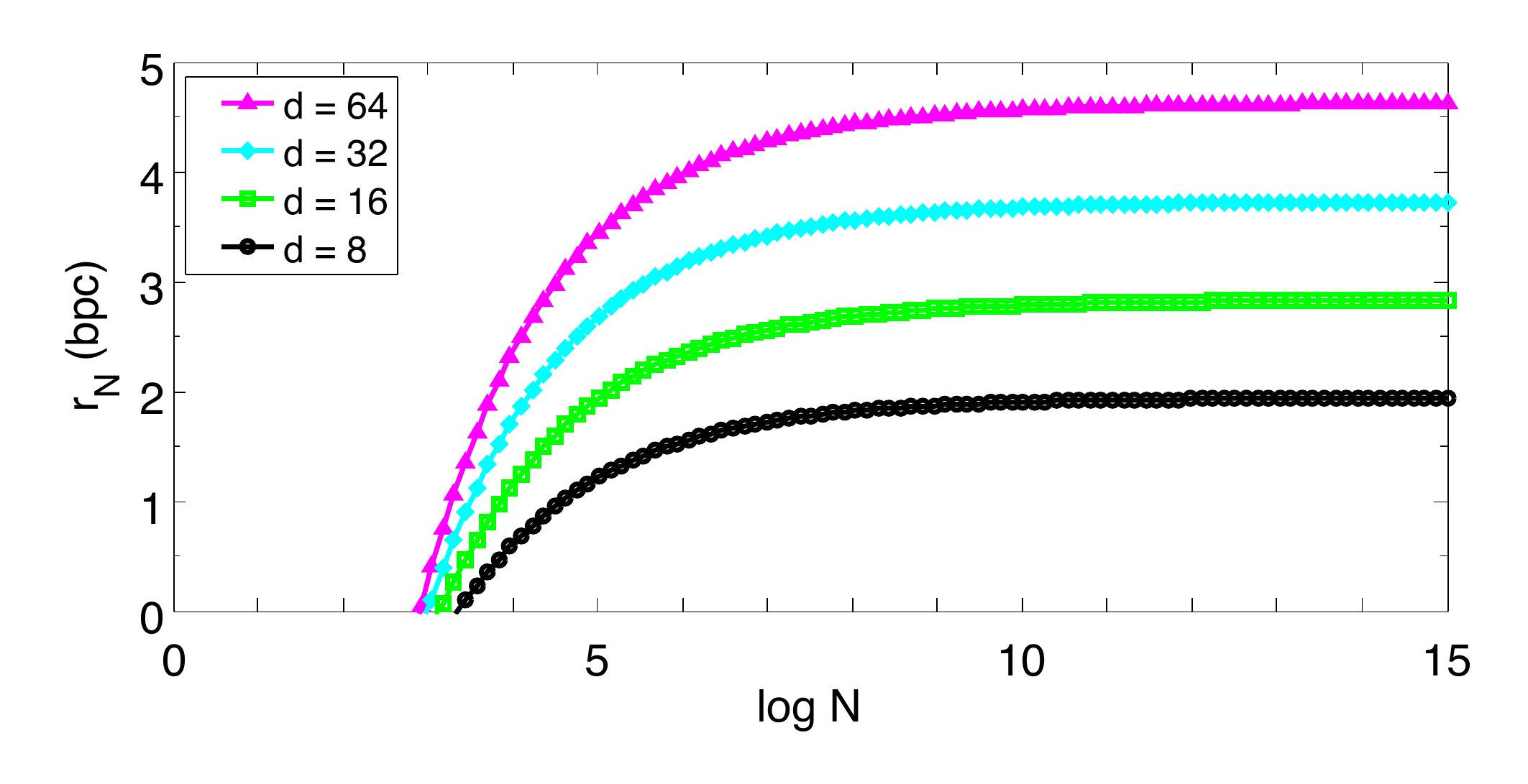}                                                                                                                                                                                                                                                                                                                                                                                                                                                                                                                                                                                                                                                                                                                                                                                                                                                                                                                                                                                                                                                                                                                                                                                                                                                                                                                                                                                                                                                                                                                                                                                                                                                                                                                                                                                                                                                                                                    
    \caption{(Color online) Plot of DO-QKD finite-key secure-key capacities in bpc (bits per frame in which Alice and Bob each detect only one event) assuming Alice and Bob observe $\hat{\sigma}_\mathrm{cor}' = 1.1\sigma_\mathrm{cor}$ and detector jitter = 2$\sigma_\mathrm{cor}$/3, where $\sigma_\mathrm{cor}$ is the correlation time. The security parameter is $\varepsilon_s = 10^{-5}$, the failure probability of the error correction is $\varepsilon_{EC} = 10^{-10}$, and the reconciliation efficiency is $\beta = 0.9$. Alice's and Bob's system detection efficiencies are 93\% \cite{NPhot.7.210}, and the dark count rate is 1000 $s^{-1}$. All other parameters were chosen to match \cite{2013.PRA.Mower.do-qkd}. 
    From top to bottom: $d = 64$, $d = 32$, $d = 16$, $d = 8$.}
    \label{mainResult}
\end{center}
\end{figure}

Figure~\ref{mainResult} plots the secure-key capacity for DO-QKD in the finite-key regime in bits per coincidence. Figure~\ref{mainResult} assumes asymmetric basis selection, zero transmission loss, estimated correlation time $\hat{\sigma}_\mathrm{cor}' = 1.1\sigma_\mathrm{cor}$, security parameter $\varepsilon_s = 10^{-5}$, and error correction code failure probability $\varepsilon_{EC} = 10^{-10}$ \cite{PRL.100.200501,PRA.82.030301}. The reconciliation efficiency is $\beta = 0.9$, which is possible using multilevel reverse reconciliation with low-density parity-check (LDPC) codes \cite{PRA.76.042305}. 
It was found that the secure-key capacity is not strongly altered by the choice of $\varepsilon_s$ \cite{PRL.100.200501,PRA.81.062343}. 
Likewise, for $d=8$ we calculated $\mathrm{max}(r_N) = 1.94$ for all security parameters between $10^{-4}$ and $10^{-7}$, and found similar results for other $d$.

An important figure of merit is the smallest $N$ at which Alice and Bob can obtain a useful amount of secure information. Figure~\ref{mainResult} shows that this occurs around $N\approx 10^4$, for the chosen parameter values. 
The inability to obtain secure key at lower $N$ values is due to the finite key length and its effect on Alice and Bob's parameter estimation. 
As $N$ gets smaller, Alice and Bob must sacrifice a larger fraction of their measurements to estimate $\xi$ to the desired accuracy. If $N$ is too small, Alice and Bob have too few measurements left to use for key generation after sacrificing the required number for parameter estimation. 

\begin{figure}
\begin{center}
    \includegraphics[width=\linewidth]{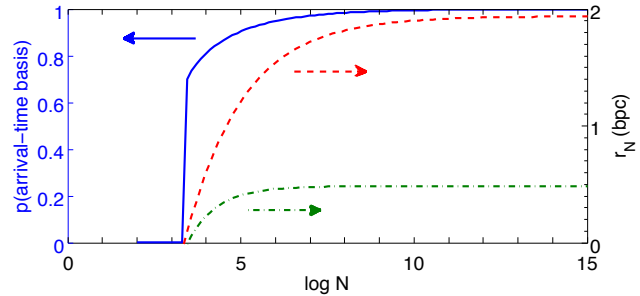}
    \caption{(Color online) Comparison of $p$ = probability of choosing the arrival-time basis (solid line, left), the secure-key capacity in bpc assuming asymmetric basis selection (dashed line, right), and the secure-key capacity in bpc assuming symmetric basis selection (dash-dotted line, right) for $d = 8$. For all $N$, the secure-key capacity is maximized by choosing $p > 1/2$. With symmetric basis selection ($p = 1/2$), the secure-key capacity is limited to 25\% of the asymptotic value.
    }
    \label{skr-p}
\end{center}
\end{figure}

The probability of choosing the arrival-time basis, $p$, directly determines the number of measurements sacrificed, $m = (1-p)^2N$. For each value of $N$, the value of $p$ is determined numerically to maximize the secure-key capacity. Figure~\ref{skr-p} plots the arrival-time basis selection probability $p$, the secure-key capacity using asymmetric basis selection, and the secure-key capacity using symmetric basis selection as functions of $N$ for $d = 8$. 
Asymmetric basis selection clearly boosts the amount of secure information per coincidence, with $p$ approaching 1 as the asymmetric secure-key capacity approaches its asymptotic value. 
In the symmetric case, where $p = 1/2$, Alice and Bob have on average only $N/2$ coincidences that were measured in the same basis: Around $N/4$ coincidences were measured in the arrival-time basis, and $N/4$ in the dispersed arrival-time basis. We continue to assume that the measurements made in the dispersed arrival-time basis are used for parameter estimation, leaving only around $n = N/4$ measurements made in the arrival-time basis for the key. With this assumption, the maximum possible secure-key capacity, even for large $N$, reaches only 25\% of the asymptotic value. 
For all $N$ that yield a positive amount of secure key, it is optimal to choose $p > 1/2$. 
However, while the asymmetric basis selection increases the secure-key capacity for all $N$ that yield a positive amount of secure key, we see numerically that it does not change the minimum $N$ required to obtain a positive amount of secure key. 

Discrete-variable QKD protocols are generally able to extract a useful amount of secure information at $N\approx 10^5$ \cite{RevModPhys.81.1301,PRL.100.200501,PRA.82.030301,NewJPhys.11.045024}. 
Continuous-variable QKD (CV-QKD) protocols require more measurements; for realistic parameter values, secure information is not obtained until $N \approx 10^8$ \cite{PRA.81.062343,NPhot.7.378}. 
Although time is a continuous variable, DO-QKD performs more like a discrete-variable protocol when considering the minimum $N$ required to obtain secure key: some secure key can be obtained even at $N \approx 10^4$.

\begin{figure}
\begin{center}
    \includegraphics[width=\linewidth]{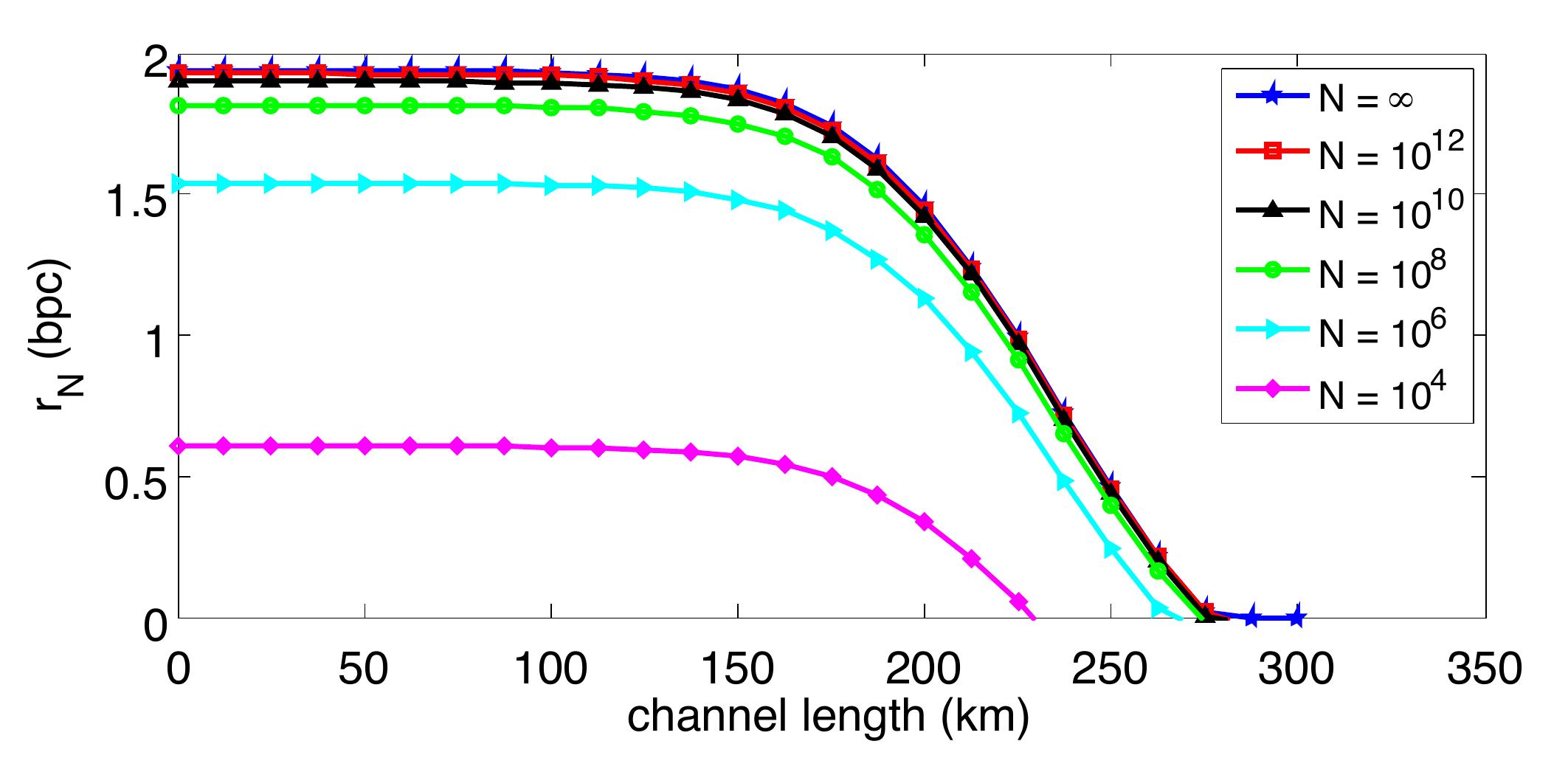}
    \caption{(Color online) Finite-key secure-key capacities in bpc versus channel length (loss) for different numbers of coincidences, $N$. $d = 8$ for all; transmission loss 0.2 dB/km; other parameters same as Figure~\ref{mainResult} and \cite{2013.PRA.Mower.do-qkd}. From top to bottom: $N = \infty$, $N = 10^{12}$, $N = 10^{10}$, $N = 10^8$, $N = 10^6$, $N = 10^4$.
    }
    \label{channel}
\end{center}
\end{figure}

We also see that even including finite-key effects, DO-QKD can reach a transmission distance $> 200$ km. This is longer than the maximum distance reached by CV-QKD protocols, which have so far seen transmission up to 80 km \cite{NPhot.7.378}. 
Figure~\ref{channel} plots the asymmetric secure-key capacity as a function of channel length for dimension $d = 8$ and various values of $N$. 

\section{Conclusion}

We have shown security against collective attacks for a high-dimensional QKD protocol in the finite-key regime. The protocol considered, DO-QKD, is robust to noise and can provide transmission of secure information at distances $> 200$ km of fiber. Working in the finite-key regime does not significantly affect the previously calculated secure-key capacity \cite{2013.PRA.Mower.do-qkd}: for experimentally achievable parameters, Alice and Bob can reach $> 90\%$ of the asymptotic secure-key capacity for a reasonable number of coincidences, $N \approx 10^{8}$, and a positive amount of secure key can be extracted after detection of as few as $N \approx 10^4$ coincidences. 

\begin{acknowledgments}
This work was supported by the DARPA Information in a Photon program, through grant W911NF-10-1-0416 from the Army Research Office, and the Columbia Optics and Quantum Electronics IGERT under NSF grant DGE-1069420. We acknowledge Greg Steinbrecher for comments on the manuscript.
\end{acknowledgments}

\bibliography{/home/cath/LaTeX/bib.bib}

\end{document}